\documentstyle[aps,12pt,epsf]{revtex} 
\begin{document} 
\pagestyle{plain} 
\setcounter{page}{1} 
\baselineskip=0.3in 
\begin{titlepage} 
\vspace{.5cm} 
 
\begin{center} 
{\Large Leading Electroweak Corrections to the Neutral Higgs  
          Boson Production at the Fermilab Tevatron } 
 
\vspace{.2in} 
 Qing Hong Cao, Chong Sheng Li \\ 
 Department of Physics, Peking University, 
Beijing 100871, P. R. China\\ 
\vspace{.2in} 
Shou Hua Zhu  \\ 
 CCAST (World Laboratory), P. O. Box 8730, Beijing 100080, P. R. China \\ 
 Institute of Theoretical Physics, Academia Sinica, P. O. Box 
 2735, Beijing 100080, P. R. China 
\end{center} 
 
\begin{footnotesize} 
\begin{center}\begin{minipage}{5in} 
\baselineskip=0.25in 
\begin{center} ABSTRACT \end{center} 
We calculate the leading electroweak corrections to the light neutral
Higgs boson production via $q\bar q'\rightarrow WH$ at the Fermilab
Tevatron in both the standard model and the minimal supersymmetric
model, which arise from the top-quark and Higgs boson loop 
diagrams. We found that the leading electroweak corrections can exceed 
the QCD corrections for favorable values of the parameters in 
the MSSM, but such corrections are only about $-1\% \sim -2\%$ 
in the SM, which are much smaller than the QCD corrections. 
For the mass region of $90 < m_{h_0} < 120$ GeV, 
the leading electroweak corrections can reach 
$-10\%$ for large $\tan\beta$, and these corrections
may be observable at a high luminosity Tevatron; 
at the least, new constraints on the $\tan\beta$ 
can be established.

\end{minipage}\end{center} 
\end{footnotesize} 
\vfill 
 
 
\end{titlepage} 
 
\newpage 
\section{Introduction} 
  One of the most important physics goals for future high energy
physics is the discovery of the Higgs boson. Recent direct search 
in the LEP2 experiments of running at $\sqrt{s}=183$ GeV via the
$e^+e^-\rightarrow Z^*H$ yields a lower bound of $\sim 89.9$ GeV on 
the Higgs mass \cite{a01}. Next year's running at $192$ GeV  
will explore up to a Higgs boson mass of about $96$ GeV 
\cite{a02}. 
After LEP2 the search for the Higgs particles will be continued
at the CERN Large Hadron Collider (LHC) for Higgs boson masses
up to the theoretical upper limit. 
Before the LHC comes into operation it is worth considering 
whether the Higgs boson can be discovered from the existing hadron 
collider, the Tevatron. Much study has been made in the detection 
of a Higgs boson at the Tevatron \cite{a03}. In Ref. \cite{a02},
it was pointed 
out that if the Higgs boson is discovered at LEP2, it should
be observed at the Tevatron's Run II with CM energy $\sqrt{s}=2$ TeV and
an integrated luminosity $\sim 10 fb^{-1}$, through the production 
subprocess $q\bar q'\rightarrow WH$, followed by $W \rightarrow \ell\bar 
\nu$ 
and $H\rightarrow b\bar b$, and if the Higgs boson lies beyond the reach
of LEP2, $m_H \geq (95-100)$ GeV, then a $5-\sigma$ discovery will be
possible in the above production sub-process in a future Run III with
an integrated luminosity $30 fb^{-1}$ for masses up to $m_H \approx
125$ GeV. Since the expected number of events is small, it is 
important to calculate the cross section as accurately as possible. 
In Ref. \cite{a05} the $ O(\alpha_s)$ QCD correction to the total 
cross section to this process have been calculated, and the QCD 
correction were found to be about $12\%$ in the $\overline{MS}$ 
scheme at the Fermilab Tevatron and the LHC in the Standard Model(SM). 
In general, the SM electroweak corrections are small comparing with the 
QCD correction. Beyond the SM, the electroweak corrections might be 
enhanced, since more Higgs bosons and the top quark 
with stronger couplings  
are involved in the loop diagrams; for example, in
the minimal supersymmtric model(MSSM)\cite{a06}, which 
predict that the lightest Higgs boson $h_0$ be less than 
$140 GeV$ \cite{a11}. 
Therefore, it is worthwhile to calculate the electroweak 
corrections to the light Higgs boson production via $q\bar 
q'\rightarrow Wh_0$. In a previous paper \cite{a07} we calculated
the $ O(\alpha_{ew} m_t^2/m_W^2)$ corrections arising from the top quark
loops to this process in both the SM and the MSSM and found that
in contrast to the QCD corrections which increases the tree-level 
cross sections, such corrections reduce the cross sections by about
$1\%\sim 2\%$ in the SM, and $1\%\sim 4\%$ in the MSSM.  
However, in addition to these top quark loops corrections, the Higgs boson
loops corrections should also be taken into account, which are of order 
$Q(\alpha_{ew} m_H^2/m_{W}^2)$, especially in the MSSM, such corrections 
may be comparable to the top quark loops corrections, and even exceed them 
for large $\tan\beta$.   
In this paper, we consider the leading electroweak
radiative corrections to the Higgs boson production at the Fermilab Tevatron 
in both the SM and the MSSM. These corrections arise from the virtual effects 
of the third family (top and bottom) of quark, neutral and charged Higgs 
bosons, and neutral and charged Goldstone bosons. We present the corrections
to production cross section versus Higgs boson mass changing in the
region of $60$ GeV $\le m_{h_0} \le 130$ GeV for different values of 
$\tan\beta$, and compare with the results of the top quark loops in the
SM and the MSSM, and also compare with the QCD corrections.

\section{Caculations} 
The leading electroweak corrections to the process
$q(p_1) \bar q'(p_2)\rightarrow W (k_1) h_0(k_2)$ 
arise from the Feynman diagrams shown in Fig 1-4. We perform 
the calculation in the 't Hooft-Feynman gauge and use dimensional 
regularization to all the ultraviolet divergence in the virtual 
loop corrections utilizing the on-mass-shell 
renormalization \cite{a08}, in which the fine-structure constant 
$\alpha$ and the physical masses are chosen to be the renormalized 
parameters, and the finite parts of the countertems are fixed by 
the renormalization conditions. As far as the parameters $\beta$ 
and $\alpha$, for the MSSM we are considering, they have to be 
renormalized, too. In the MSSM they are not independent. 
Nevertheless, we follow the approach of Mendez and Pomarol \cite{a09} 
in which they consider them as independent renormalized parameters 
and fixed the corresponding renormalization constant by a 
renormalization condition that the on-mass-shell $H^+\bar \ell\nu 
_{\ell}$ and $h_0\bar \ell\ell$ couplings keep the forms of Eq.(3) 
of Ref. \cite{a09} to all order of perturbation theory. 
 
  We define the Mandelstam variables as 
\begin{eqnarray} 
\hat{s}&=&(p_1+p_2)^2=(k_1+k_2)^2\nonumber \\ 
\hat{t}&=&(p_1-k_1)^2=(p_2-k_2)^2 \nonumber \\ 
\hat{u}&=&(p_1-k_2)^2=(p_2-k_1)^2. 
\end{eqnarray} 
 
 The relevant renormalization constants are defined as 
\begin{eqnarray} 
m^2_{W0}=m^2_W+ \delta m^2_W, \ \ \ \  m^2_{Z0}=m^2_Z+ \delta m^2_Z, 
\end{eqnarray} 
\begin{eqnarray} 
\tan \beta_0=(1+\delta Z_{\beta})\tan \beta,\ \ \ \ 
\sin \alpha_0=(1+\delta Z_{\alpha})\sin \alpha, 
\end{eqnarray} 
\begin{eqnarray} 
W^{\pm \mu}_0=Z^{1/2}_WW^{\pm\mu}+ 
i Z^{1/2}_{H^\pm W}\partial^{\mu}H^{\pm},\ \ \ \ 
H^{\pm}_0=(1+\delta Z_{H^{\pm}})^{1/2}H^{\pm}, 
\end{eqnarray} 
\begin{eqnarray} 
h_0=(1+\delta h_0)^{1/2}h+ 
Z^{1/2}_{h_0H}H,\ \ \ \ 
H_0=(1+\delta Z_H)^{1/2}H+Z^{1/2}_{Hh_0}h 
\end{eqnarray}

 Taking into account the leading electroweak corrections, the renormalized 
amplitude for $q\bar q'\rightarrow Wh_0$  can be written as 
\begin{eqnarray} 
M_{ren}=M_0+\delta M^{self}+\delta M^{vertex}, 
\end{eqnarray} 
where $M_0$ is the amplitude at the tree level, $\delta M^{self}$ 
and $\delta M^{vertex}$ represent the corrections arising 
from the self-energy and vertex  diagrams, respectively. $M_0$ is given by 
\begin{eqnarray} 
M_0= {e^2 m_W \sin(\beta-\alpha) 
\over \sqrt{2} (m_W^2-\hat{s}) \sin\theta_w^2} 
\bar{d}(p_2)\rlap/\epsilon P_Lu(p_1), 
\end{eqnarray} 
where $P_{L,R}\equiv(1\mp\gamma_5)/2$. 
$\delta M^{self}$ is given by 
\begin{eqnarray} 
\delta M^{self}= {\delta m_W^2 +( m_W^2 -\hat{s}) \delta Z_W 
\over \hat{s} - m_W^2} M_0+\delta M^{self}_t+\delta M^{self}_H 
\end{eqnarray} 
$\delta M^{self}_t$ represents the correction arising from the 
top-quark self-energy diagrams to the order $O(\alpha_{ew} 
m_t^2/m_W^2)$, $\delta M^{self}_H$ represents the corrections 
arising from the Higgs boson self-energy diagrams to the order 
$O(\alpha_{ew} m_H^2/m_W^2)$. $H=M_H^{\pm},h,M_H^0,A$. 
The renormalization constant 
$\delta m_W^2$ and $\delta Z_{W}$ are presented in the appendix-A . 
$\delta M^{self}_t$ and $\delta M^{self}_H$ are presented in the 
Appendix-B.

$\delta M^{vertex}$ is given by 
\begin{eqnarray} 
\delta M^{vertex}&=& M_0 \left[ {1\over 2} \delta Z_{h_0}+{\delta m_W^2 -\delta m_Z^2 
\over 2 (m_Z^2-m_W^2)} +{\delta m_Z^2 \over m_Z^2}+{\delta m_W^2 \over m_W^2}+{\delta e \over e} 
\right. 
\nonumber\\ 
&+& 
\left. 
\cot (\beta-\alpha)(Z_{Hh_0}^{1/2}+\sin\beta \cos\beta \delta Z_\beta - 
\tan\alpha \delta Z_\alpha)\right] \nonumber \\ 
&+&f_1^{vertex} \bar{d}(p_2)\rlap/\epsilon P_L u(p_1)\nonumber \\ 
&+&f_2^{vertex} \bar{d}(p_2)\rlap/p_1 P_L u(p_1) \epsilon.p_1\nonumber \\ 
&+&f_3^{vertex} \bar{d}(p_2)\rlap/p_1 P_L u(p_1) \epsilon.p_2, 
\end{eqnarray} 
where\\ 
\begin{eqnarray} 
\delta Z_\beta &=& {\delta m_Z^2 -\delta m_W^2 
\over 2 (m_Z^2-m_W^2)} 
-{\delta m_Z^2 \over 2 m_Z^2}+{\delta m_W^2 \over 2 m_W^2}- 
{1\over 2} \delta Z_{H^\pm} -{m_W\over \tan\beta} Z_{WH^\pm}^{1/2}, 
\end{eqnarray} 
\begin{eqnarray} 
\delta Z_\alpha=&-&\sin^2\beta \delta Z_\beta +{\delta m_Z^2 -\delta m_W^2 
\over 2 (m_Z^2-m_W^2)} 
-{\delta m_Z^2 \over 2 m_Z^2}+{\delta m_W^2 \over 2 m_W^2}\nonumber\\ 
&-&{1\over 2}\delta 
Z_{h_0}+{\cos\alpha\over\sin\alpha}Z_{Hh_0}^{1/2} , 
\end{eqnarray} 
 
Note that $\delta e/e$ appearing in Eq. (9) contains no $O(\alpha m_t^2/m_W^2)$
or $O(\alpha m_H^2/m_W^2)$ terms and need not be considered in our 
calculations. Although the renormalization constant $Z_{Hh_0}^{1/2}$ 
appears in $\delta Z_\alpha$ and the vertex counterterm, they cancel 
each other with accuracy and give no contribution to our 
calculation. Others renormalization constants are listed in the 
Appendix-A and the vertex factors $f_{1,2,3}^{vertex}$ are given in 
the Appendix-C.

 The corresponding amplitude squared for the process 
 $q \bar q' 
\rightarrow W h_0 $ 
 can be written as 
\begin{eqnarray} 
\overline{{\sum}}\left| M_{ren}\right|^2=\overline{{\sum}}\left| M_{0}\right|^2 
 +2 Re \overline{{\sum}}(\delta M^{self}+ \delta M^{vertex}) M_0^\dagger, 
\end{eqnarray} 
where the bar over the summation recalls average over initial partons spins. 
The cross section of  process $q \bar q' 
\rightarrow W h_0 $ is 
\begin{eqnarray} 
\hat{\sigma} =\int_{\hat{t}_{min}}^{\hat{t}_{max} }{1\over 16\pi \hat{s }^2}\bar 
{ \sum_{ spins} }\left|  M\right|^2 d \hat{t} 
\end{eqnarray} 
with 
\begin{eqnarray} 
\hat{t}_{min}&=&{m_{h_0}^2+m_W^2-\hat{s}\over 2}-\sqrt{ 
   (\hat{s}-(m_{h_0}+m_W)^2)(\hat{s}-(m_{h_0}-m_W)^2)/2} \nonumber \\ 
\hat{t}_{max}&=&{m_{h_0}^2+m_W^2-\hat{s}\over 2}+\sqrt{ 
   (\hat{s}-(m_{h_0}+m_W)^2)(\hat{s}-(m_{h_0}-m_W)^2)/2}. 
\end{eqnarray} 
The total cross section of $P\bar{P}\rightarrow q \bar q' \rightarrow 
W h_0$ 
can be obtained by folding the $\hat{\sigma}$  with the 
parton luminosity 
\begin{eqnarray} 
\sigma (s)=\int^{1}_{(m_{h_0}+m_W)/\sqrt{s}} dz 
{dL\over dz} \hat{\sigma} (q \bar q'\rightarrow W h_0 \mbox{ at 
$\hat{s}=z^2 s$}), 
\end{eqnarray} 
where $\sqrt{s}$ and $\sqrt{\hat{s}}$ is the CM energy of $P\bar{P}$ and 
$q \bar q'$, respectively, and $dL/dz$ is the parton 
luminosity, which is defined as 
\begin{eqnarray} 
{dL\over dz}=2 z \int_{z^2}^1{dx\over x} 
f_{q /P}(x,q^2) f_{q' /\bar{P}}(z^2/x,q^2), 
\end{eqnarray} 
where $f_{q/P}(x,q^2)$ and $f_{q'/\bar{P}}(z^2/x,q^2)$ are 
the parton distribution function \cite{CTEQ}.

\section{Numerical Results} 
In our numerical calculations, the SM parameters were taken to be 
$m_W=80.33 GeV$, $m_Z=91.187 GeV$, $m_t=176 GeV$, $m_b=4.5 GeV$ and 
$\alpha_{ew}={1\over 128}$. Moreover, we use the relation \cite{add} 
between the Higgs boson masses $m_{h_0,H,A,H^{\pm}}$ and parameters 
$\alpha, \beta$ at one-loop, and choose $m_{h_0}$ and $\tan \beta$ 
as two independent input parameters. As explained in 
Ref. \cite{a09}, there is a small inconsistency in doing so since 
the parameters $\alpha$ and $\beta$ of Ref. \cite{a11} are not the 
ones defined by the conditions given by Eq.(3) of Ref. \cite{a09}. 
Nevertheless, this difference would only induce a higher order 
change \cite{a09}. We will limit the value of $\tan \beta$ to be in 
the range $2\leq \tan \beta\leq 30$, which are consistent with 
ones required by the most popular MSSM model with scenarios 
motivated by current low energy data (including $\alpha_s$, $R_b$ 
and the branching ratio of $b\rightarrow s\gamma$). 

 In Figs. 5 and 6 we present both the top quark loop corrections \cite{a07} 
 and the leading electroweak radiative corrections as a function of $m_h$
for the different values of $\tan\beta$ using the CTEQ3L parton
distributions for the tree-level cross sections $\sigma_0$ and 
the CTEQ3M parton distributions for the corrections 
$\delta \sigma$. From Fig. 5 one sees that the leading electroweak 
corrections are almost the same as the top quark loop corrections in the SM,
which means that the corrections arising from the Higgs boson loops are
negligibly small in the SM. And these corrections are not sensitive to
the mass of the Higgs boson and amount to $1\% \sim 2\%$ reduction in the
tree-level total cross sections.  
However, in the MSSM, the Higgs boson loops corrections are important,
and especially for large $\tan\beta (>4)$, they can exceed the top quark loop 
corrections. As a result, the leading electroweak corrections are much larger than
the top quark loops corrections. As shown in Fig. 6, for $\tan\beta=30$,
the leading electroweak corrections decrease the cross section by 
$35\%$
when $m_h=60$ GeV, while the top quark loops corrections decrease only 
about $-4\%$. However, these corrections are sensitive to $m_{h_0}$. 
For $m_{h_0}$ 
in the range $90-120$ GeV, the leading electroweak corrections and the top
quark loops corrections drop to about $-10\% \sim -2\%$ and $-3\% \sim -1\%$, 
respectively, which indicate that the leading electroweak corrections are 
still obviously larger than one from the top quark loops. Only in the 
vicinity of $m_{h_0} \approx 130$ GeV for all values of $\tan\beta$  
are the leading electroweak and top loop corrections about the same 
as one in the SM.   
  
Comparing with the QCD corrections to this process \cite{a05}, which 
increase the tree-level total cross sections by about $12\%$, 
we find that in general the leading electroweak corrections partly cancel 
the QCD corrections, but for large $\tan\beta$, the magnitude of 
the former can even exceed the latter, and have to be considered in
searching the light Higgs boson in the MSSM through this process at the
Fermilab Tevatron.

  To summarize, the leading electroweak radiative corrections, 
which combine the top quark and the Higgs boson loops contributions, 
can exceed the QCD corrections for favorable values of the parameters in 
the MSSM, but such corrections are only about $-1\% \sim -2\%$ 
in the SM, which are much smaller than the QCD corrections. 
The mass region of $90 < m_{h_0} < 120$ GeV 
is the interesting window for searching the Higgs boson 
at the Tevatron, in which the leading electroweak corrections vary from
$-10\%$ to $-2\%$ for $\tan\beta = 30$, and these corrections
may be observable at a high luminosity Tevatron; 
at the least, new constraints on the $\tan\beta$ 
can be established.

\section*{Acknowledgments} 
This work was supported in part by the National Natural Science Foundation 
of China and a grant from the State Commission of Science and 
Technology of China. 

 
\section* {Appendix-A: the renormalization constants} 
\begin{eqnarray} 
\delta m_W^2&=& 
{N_ce^2m_t^2\over 96\pi ^2\sin^2\theta_W} 
\left[-2+{m_t^2\over m_W^2}[B_0(0,m_b^2,m_t^2)-B_0(m_W^2,m_b^2,m_t^2)]\right.\nonumber\\ 
&+&\left.2B_0(0,m_b^2,m_t^2)-B_0(m_W^2,m_b^2,m_t^2)-4B_0(0,m_t^2,m_t^2)\right]\nonumber\\ 
&+&{e^2\sin^2 (\beta - \alpha )\over 192 \pi ^2 \sin^2\theta _W } 
\left[-2 m_H^2-4 {m_{H^{+}}}^2\right.\nonumber\\ 
&+&{m_H^4\over m_W^2} 
[B_0(m_W^2,m_H^2,m_{H^+}^2)-B_0(0,m_H^2,{m_{H^{+}}}^2)]\nonumber\\ 
&+&{2m_H^2{m_{H^{+}}}^2\over m_W^2} 
[B_0(0,m_H^2,{m_{H^{+}}}^2)-B_0(m_W^2,m_H^2,{m_{H^{+}}}^2)]\nonumber\\ 
&+&{{m_{H^{+}}}^4\over m_W^2} 
[B_0(m_W^2,m_H^2,{m_{H^{+}}}^2)-B_0(0,m_H^2,{m_{H^{+}}}^2)]\nonumber\\ 
&+&m_H^2[-2B_0(m_W^2,m_H^2,{m_{H^{+}}}^2)-B_0(0,m_H^2,{m_{H^{+}}}^2)]\nonumber\\ 
&+&\left.{m_{H^{+}}}^2[-2B_0(m_W^2,m_H^2,{m_{H^{+}}}^2)+B_0(0,m_H^2,{m_{H^{+}}}^2) 
-2B_0(0,m_{H^+}^2,{m_{H^{+}}}^2)]\right]\nonumber\\ 
&+&{e^2\cos^2(\beta - \alpha)\over 192 \pi ^2 \sin^2\theta _W } 
\left[-2{m_{h_{0}}}^2-4 {m_{H^{+}}}^2\right.\nonumber\\ 
&+&{{m_{H^{+}}}^4\over m_W^2} 
[B_0(m_W^2,{m_{H^{+}}}^2,{m_{h_{0}}}^2)-B_0(0,{m_{H^{+}}}^2,{m_{h_{0}}}^2)]\nonumber\\ 
&+&{2{m_{h_{0}}}^2{m_{H^{+}}}^2\over m_W^2} 
[B_0(0,{m_{H^{+}}}^2,{m_{h_{0}}}^2)-B_0(m_W^2,{m_{H^{+}}}^2,{m_{h_{0}}}^2)]\nonumber\\ 
&+&{{m_{h_{0}}}^4\over m_W^2} 
[B_0(m_W^2,{m_{H^{+}}}^2,{m_{h_{0}}}^2)-B_0(0,{m_{H^{+}}}^2,{m_{h_{0}}}^2)]\nonumber\\ 
&+&{m_{H^{+}}}^2[-2B_0(m_W^2,{m_{H^{+}}}^2,{m_{h_{0}}}^2) 
-2B_0(0,{m_{H^{+}}}^2,{m_{H^{+}}}^2)+B_0(0,{m_{H^{+}}}^2,{m_{h_{0}}}^2)]\nonumber\\ 
&+&\left. {m_{h_{0}}}^2[-B_0(0,{m_{H^{+}}}^2,{m_{h_{0}}}^2) 
-2B_0(m_W^2,{m_{H^{+}}}^2,{m_{h_{0}}}^2)]\right]\nonumber\\ 
&+&{e^2\over 192 \pi ^2 \sin^2\theta _W } \left[-2{m_{H^{+}}}^2-4 
m_A^2\right.\nonumber\\ 
&+&{m_A^4 \over m_W^2} 
[B_0(m_W^2,m_A^2,{m_{H^{+}}}^2) 
-B_0(0,m_A^2,{m_{H^{+}}}^2)]\nonumber\\ 
&+& {2m_A^2 {m_{H^{+}}}^2\over m_W^2} [B_0(0,m_A^2,{m_{H^{+}}}^2) 
 -B_0(m_W^2,m_A^2,{m_{H^{+}}}^2)]\nonumber\\ 
&+& {{m_{H^{+}}}^4\over m_W^2} [B_0(m_W^2,m_A^2,{m_{H^{+}}}^2) 
-B_0(0,m_A^2,{m_{H^{+}}}^2)]\nonumber\\ 
&+& m_A^2[-2B_0(m_W^2,m_A^2,{m_{h^{+}}}^2) 
-2B_0(0,m_A^2,m_A^2) 
+B_0(0,m_A^2,{m_{H^{+}}}^2)]\nonumber\\ 
&+&\left. 
{m_{H^{+}}}^2[-B_0(0,m_A^2,{m_{H^{+}}}^2) 
-2B_0(m_W^2,m_A^2,{m_{H^{+}}}^2)]\right]\nonumber\\ 
&+&{e^2\cos^2(\beta - \alpha )\over 192 \pi ^2 \sin ^2\theta _W } 
\left[-4 m_H^2+ 
{m_H^4\over m_W^2} 
[B_0(m_W^2,m_H^2,m_W^2)-B_0(0,m_H^2,m_W^2)]\right.\nonumber\\ 
&+&\left. m_H^2[-4B_0(m_W^2,m_H^2,m_W^2)-2B_0(0,m_H^2,m_H^2) 
+3B_0(0,m_H^2,m_W^2)]\right]\nonumber\\ 
&+&{e^2\sin ^2(\beta 
-\alpha )\over 192 \pi ^2 \sin ^2\theta _W } 
\left[-4 {m_{h_{0}}}^2+ 
{{m_{h_{0}}}^4\over m_W^2} 
[B_0(m_W^2,{m_{h_{0}}}^2,m_W^2)-B_0(0,{m_{h_{0}}}^2,m_W^2)]\right.\nonumber\\ 
&+&\left. 
{m_{h_{0}}}^2[-4B_0(m_W^2,{m_{h_{0}}}^2,m_W^2)-2B_0(0,{m_{h_{0}}}^2,{m_{h_{0}}}^2) 
+3B_0(0,{m_{h_{0}}}^2,m_W^2)]\right]\nonumber\\
&+&{e^2 m_H^2\over 
64 \pi ^2\sin^2\theta_W} [1+B_0(0,m_H^2,m_H^2)]\nonumber\\ 
&+&{e^2{m_{H^{+}}}^2\over 32\pi^2\sin ^2\theta_W} 
[1+B_0(0,{m_{H^{+}}}^2,{m_{H^{+}}}^2)]\nonumber\\ &+&{e^2 
m_A^2\over64 \pi^2\sin^2\theta_W} [1+B_0(0,m_A^2,m_A^2)]\nonumber\\ 
&+&{e^2 {m_{h_{0}}}^2\over 64\pi^2\sin ^2\theta _W 
}[1+B_0(0,{m_{h_{0}}}^2,{m_{h_{0}}}^2)] 
\end{eqnarray}

\begin{eqnarray} 
\delta m_Z^2 
&=&{N_Ce^2 m_t^2\over 288 \cos^2\theta_W 
\pi^2 
\sin^2\theta_W}\left[-18B_0(0,m_t^2,m_t^2) 
+48 \sin(\theta_W)^2 B_0(0,m_t^2,m_t^2)\right.\nonumber\\ 
&-&64\sin^4\theta_W 
B_0(0,m_t^2,m_t^2)-9B_0(m_Z^2,m_t^2,m_t^2)\nonumber\\ 
&-&\left.48\sin^2\theta_WB_0(m_Z^2,m_t^2,m_t^2) 
+64\sin^2\theta_WB_0(0,m_t^2,m_t^2)\right]\nonumber\\ 
&+&{e^2(\sin^2\theta_W-\cos^2\theta_W)^2\over 96 
\cos^2\theta_W \pi^2 
\sin^2\theta_W}\left[-3{m_{H^{+}}}^2 
-{m_{H^{+}}}^2B_0(0,{m_{H^{+}}}^2,{m_{H^{+}}}^2)\right.\nonumber\\ 
&-&\left.2{m_{H^{+}}}^2B_0(m_Z^2,{m_{H^{+}}}^2,{m_{H^{+}}}^2) 
\right]\nonumber\\ 
&+&{e^2 
\cos^2(\alpha-\beta)\over 192\cos^2\theta_W\pi^2\sin^2\theta_W} 
\left[-2m_H^2+{m_H^4\over m_Z^2}[B_0(m_Z^2,m_H^2,m_Z^2)-B_0(0,m_H^2,m_Z^2)] 
\right.\nonumber\\ 
&+&\left.m_H^2[B_0(0,m_H^2,m_Z^2) 
-4B_0(m_Z^2,m_H^2,m_Z^2)]\right]\nonumber\\ 
&+&{e^2 
\sin^2(\alpha-\beta)\over 192\cos^2\theta_W\pi^2\sin^2\theta_W} 
\left[-2m_H^2-4m_A^2\right.\nonumber\\ 
&+&{(m_H^2-m_A^2)^2\over m_Z^2} 
[B_0(m_Z^2,m_A^2,m_H^2)-B_0(0,m_A^2,m_H^2)]\nonumber\\ 
&+&m_H^2[-B_0(0,m_A^2,m_H^2) 
-2B_0(m_Z^2,m_A^2,m_H^2)]\nonumber\\ 
&+&\left.m_A^2[-2B_0(m_Z^2,m_A^2,m_H^2) 
-2B_0(0,m_A^2,m_A^2) 
+B_0(0,m_A^2,m_H^2)]\right]\nonumber\\ 
&+&{e^2 
\sin^2(\alpha-\beta)\over 192\cos^2\theta_W\pi^2\sin^2\theta_W} 
\left[-2{m_{h_{0}}}^2+{m_{h_0}^4 \over m_Z^2} 
[B_0(m_Z^2,{m_{h_{0}}}^2,m_Z^2)-B_0(0,{m_{h_{0}}}^2,m_Z^2)]\right.\nonumber\\ 
&+&\left.{m_{h_{0}}}^2[B_0(0,{m_{h_{0}}}^2,m_Z^2)-4B_0(m_z^2,{m_{h_{0}}}^2,m_Z^2)] 
\right]\nonumber\\ 
&+&{e^2 
\cos^2(\alpha-\beta)\over 192\cos^2\theta_W\pi^2\sin^2\theta_W} 
\left[-2{m_{h_{0}}}^2-4m_A^2\right.\nonumber\\ 
&+&{({m_{h_{0}}}^2-m_A^2)^2\over m_Z^2} 
[B_0(m_Z^2,m_A^2,{m_{h_{0}}}^2)-B_0(0,m_A^2,{m_{h_{0}}}^2)]\nonumber\\ 
&+&{m_{h_{0}}}^2[-B_0(0,m_A^2,{m_{h_{0}}}^2) 
-2B_0(m_Z^2,m_A^2,{m_{h_{0}}}^2)]\nonumber\\ 
&+&\left.m_A^2[-2B_0(m_Z^2,m_A^2,{m_{h_{0}}}^2) 
-2B_0(0,m_A^2,m_A^2) 
+B_0(0,m_A^2,{m_{h_{0}}}^2)]\right]\nonumber\\ 
&+&{e^2(\sin^2\theta_W-\cos^2\theta_W)^2\over 32 
\cos^2\theta_W\pi^2\sin^2\theta_W}{m_{H^{+}}}^2 
\left[1+B_0(0,{m_{H^{+}}}^2,{m_{H^{+}}}^2)\right]\nonumber\\ 
&+&{e^2\over 64 
\cos^2\theta_W \pi^2\sin ^2\theta_W}m_H^2\left[1+B_0(0,m_H^2,m_H^2)\right]\nonumber\\ 
&+&{e^2\over 64 
\cos^2\theta_W \pi^2\sin^2\theta_W}{m_{h_{0}}}^2 
\left[1+B_0(0,{m_{h_{0}}}^2,{m_{h_{0}}}^2)\right]\nonumber\\ 
&+&{e^2\over 64 
\cos^2\theta_W\pi^2\sin^2\theta_W}m_A^2 
\left[1+B_0(0,m_A^2,m_A^2)\right]\nonumber\\ 
\end{eqnarray} 
 
\begin{eqnarray} 
\delta Z_W &=&
{N_Ce^2\over 288\pi^2m_W^4\sin^2\theta_W} 
\left[ 
2m_W^4-6m_W^4B_0(m_W^2,m_b^2,m_t^2)\right.\nonumber\\ 
&+&3m_t^4[B_0(m_W^2,m_b^2,m_t^2)-B_0(m_W^2,m_b^2,m_t^2)]\nonumber\\ 
&+&\left.3m_W^2(m_t^4+m_t^2m_W^2-2m_W^4)G_0(m_W^2,m_b^2,m_t^2)\right]\nonumber\\ 
&+&{e^2\sin^2 (\beta - \alpha )\over 576 m_W^4\pi ^2\sin^2\theta_W} 
\left[-2m_W^4-3m_W^4B_0(m_W^2,m_H^2,{m_{H^{+}}}^2)\right.\nonumber\\ 
&+&3(m_H^2-{m_{H^{+}}}^2)^2[-B_0(0,m_H^2,{m_{H^{+}}}^2) 
+B_0(m_W^2,m_H^2,{m_{H^{+}}}^2)]\nonumber\\ 
&+&[6m_H^2m_W^2{m_{H^{+}}}^2-3m_H^4m_W^2-3{m_{H^{+}}}^4m_W^2\nonumber\\ 
&+&\left.6m_H^2m_W^4+6{m_{H^{+}}}^2m_W^4-3m_W^6]G_0(m_W^2,m_H^2,{m_{H^{+}}}^2) 
\right]\nonumber\\ 
&+&{e^2\cos^2(\beta - \alpha )\over 576 m_W^4\pi ^2 \sin 
^2\theta_W } 
\left[ 
-2m_W^4-3m_W^4B_0(m_W^2,{m_{H^{+}}}^2,{m_{h_{0}}}^2)\right.\nonumber\\ 
&+&3({m_{h_{0}}}^2-{m_{H^{+}}}^2)^2[B_0(m_W^2,{m_{H^{+}}}^2,{m_{h_{0}}}^2) 
-B_0(0,{m_{H^{+}}}^2,{m_{h_{0}}}^2)]\nonumber\\ 
&+&[6{m_{h_{0}}}^2m_W^2{m_{H^{+}}}^2-3{m_{h_{0}}}^4m_W^2-3{m_{H^{+}}}^4m_W^2\nonumber\\ 
&+&\left.6{m_{h_{0}}}^2m_W^4+6{m_{H^{+}}}^2m_W^4-3m_W^6] 
G_0(m_W^2,{m_{H^{+}}}^2,{m_{h_{0}}}^2) 
\right]\nonumber\\ 
&+&{e^2\over 576 m_W^4\pi ^2 \sin 
^2\theta_W} 
\left[ 
-2m_W^4-3m_W^4B_0(m_W^2,m_A^2,{m_{H^{+}}}^2)\right.\nonumber\\ 
&+&3(m_A^2-{m_{H^{+}}}^2)^2[-B_0(0,m_A^2,{m_{H^{+}}}^2) 
+B_0(m_W^2,m_A^2,{m_{H^{+}}}^2)]\nonumber\\ 
&+&[6m_A^2m_W^2{m_{H^{+}}}^2-3m_A^4m_W^2-3{m_{H^{+}}}^4m_W^2\nonumber\\ 
&+&\left.6m_A^2m_W^4+6{m_{H^{+}}}^2m_W^4-3m_W^6]G_0(m_W^2,m_A^2,{m_{H^{+}}}^2) 
\right]\nonumber\\ 
&+&{e^2\cos^2(\beta - \alpha )\over 576 m_W^4\pi ^2 \sin 
^2\theta_W} 
\left[-2m_W^2-3m_W^4B_0(0,m_H^2,m_W^2)\right.\nonumber\\ 
&+&3m_H^4[B_0(m_W^2,m_H^2,m_W^2)-B_0(0,m_H^2,m_W^2)]\nonumber\\ 
&+&6m_H^2m_W^2[B_0(0,m_H^2,m_W^2)-B_0(m_W^2,m_H^2,m_W^2)]\nonumber\\ 
&+&\left.(12m_H^2m_W^4-3m_H^4m_W^2)G_0(m_W^2,m_H^2,m_W^2)\right]\nonumber\\ 
&+&{e^2\sin^2 (\beta - \alpha )\over 576 m_W^4\pi ^2 \sin 
^2\theta_W} 
\left[-2m_W^2-3m_W^4B_0(0,{m_{h_{0}}}^2,m_W^2)\right.\nonumber\\ 
&+&3{m_{h_{0}}}^4[B_0(m_W^2,{m_{h_{0}}}^2,m_W^2)-B_0(0,{m_{h_{0}}}^2,m_W^2)]\nonumber\\ 
&+&6{m_{h_{0}}}^2m_W^2[B_0(0,{m_{h_{0}}}^2,m_W^2)-B_0(m_W^2,{m_{h_{0}}}^2,m_W^2)]\nonumber\\ 
&+&\left.(12{m_{h_{0}}}^2m_W^4-3{m_{h_{0}}}^4m_W^2)G_0(m_W^2,{m_{h_{0}}}^2,m_W^2) 
\right]\nonumber\\ 
\end{eqnarray} 
 
\begin{eqnarray} 
\delta {Z_{H^{\pm}W}}^{1/2}&=&
{N_Ce^2m_t^2\cot\beta\over32{m_{H^{+}}}^2m_W^3\pi^2\sin^2\theta_W} 
\left[ 
m_t^2(B_0(0,m_b^2,m_t^2)-B_0({m_{H^{+}}}^2,m_b^2,m_t^2))\right.\nonumber\\ 
&+&\left.{m_{H^{+}}}^2B_0({m_{H^{+}}}^2,m_b^2,m_t^2)\right]\nonumber\\ 
&-&{e^2\sin (\beta - \alpha )({m_{H^{+}}}^2-m_H^2)\over 64 
m_W^2{m_{H^{+}}}^2 \pi^2 
\sin ^2\theta_W \cos \theta_W } 
[ 2\cos\theta_W 
m_W\cos(\alpha-\beta)-m_Z\cos2\beta\cos(\alpha+\beta)]\nonumber\\ 
&\times&[B_0({m_{H^{+}}}^2,m_H^2,{m_{H^{+}}}^2)-B_0(0,m_H^2,{m_{H^{+}}}^2)]\nonumber\\ 
&+&{e^2\sin (\beta + \alpha )({m_{h_{0}}}^2-{m_{H^{+}}}^2)\over 64 
m_W^2{m_{H^{+}}}^2 \pi^2 
\sin ^2\theta_W \cos \theta_W } 
[2\cos\theta_W 
m_W\sin(\alpha-\beta)-m_Z\cos2\beta\sin(\alpha+\beta)]\nonumber\\ 
&\times&[B_0({m_{H^{+}}}^2,{m_{H^{+}}}^2,{m_{h_{0}}}^2)-B_0(0,{m_{H^{+}}}^2,{m_{h_{0}}}^2)]\nonumber\\ 
&+&{e^2\sin 2\beta(m_H^2-m_W^2)\over 64 m_W^2{m_{H^{+}}}^2 \pi^2 
\sin^2 \theta_W \cos \theta_W } 
[\cos\theta_W 
m_W\sin(\alpha-\beta)+m_Z\cos(\beta+\alpha)\sin(2\beta)]\nonumber\\ 
&\times&[-B_0({m_{H^{+}}}^2,m_H^2,m_W^2)+B_0(0,m_H^2,m_W^2)]\nonumber\\ 
&+&{e^2\sin (\alpha-\beta)({m_{h_{0}}}^2-m_W^2)\over 64 
m_W^2{m_{H^{+}}}^2 \pi^2 
\sin ^2\theta_W\cos \theta_W} 
[-\cos\theta_Wm_W\cos(\alpha-\beta)+m_Z\sin(\beta+\alpha)\sin(2\beta)]\nonumber\\ 
&\times&[-B_0({m_{H^{+}}}^2,{m_{h_{0}}}^2,m_W^2)+B_0(0,{m_{h_{0}}}^2,m_W^2)]\nonumber\\ 
&+&{e^2\sin (\alpha-\beta)\cos(\alpha-\beta)\over 
64m_W{m_{H^{+}}}^2\pi^2\sin^2\theta_W } 
\left[m_H^2[B_0(0,m_H^2,m_W^2)-B_0({m_{H^{+}}}^2,m_H^2,m_W^2)]\right.\nonumber\\ 
&-&\left.3{m_{H^{+}}}^2B_0({m_{H^{+}}}^2,m_H^2,m_W^2)\right]\nonumber\\ 
&+&{e^2\sin (\alpha-\beta)\cos(\alpha-\beta)\over64m_W{m_{H^{+}}}^2 
\pi^2\sin ^2\theta_W} 
\left[{m_{h_{0}}}^2[-B_0(0,{m_{h_{0}}}^2,m_W^2)+B_0({m_{H^{+}}}^2,{m_{h_{0}}}^2,m_W^2)] 
\right.\nonumber\\ 
&+&\left.3{m_{H^{+}}}^2B_0({m_{H^{+}}}^2,{m_{h_{0}}}^2,m_W^2)\right]\nonumber\\ 
\end{eqnarray} 
 
\begin{eqnarray} 
\delta Z_{H^{\pm}}&=&
{N_Ce^2m_t^2\cot(\beta)^2\over 32m_W^2\pi^2\sin^2\theta_W} 
[-B_0({m_{H^{+}}}^2,m_b^2,m_t^2)+(m_t^2-{m_{H^{+}}}^2) 
G_0({m_{H^{+}}}^2,m_W^2,m_t^2)]\nonumber\\ 
&+&{e^2(\sin^2\theta_W-\cos^2\theta_W)^2{m_{H^{+}}}^2\over16\pi^2\sin^2\theta_W\cos^2\theta_W} 
G_0({m_{H^{+}}}^2,{m_{H^{+}}}^2,m_Z^2)\nonumber\\ 
&+&{e^2\sin(\alpha-\beta)^2({m_{H^{+}}}^2+m_H^2)\over32\pi^2\sin^2\theta_W} 
G_0({m_{H^{+}}}^2,m_H^2,m_W^2)\nonumber\\ 
&+&{e^2\cos^2(\alpha-\beta)({m_{H_{0}}}^2+{m_{H^{+}}}^2)\over32\pi^2\sin^2\theta_W} 
G_0({m_{H^{+}}}^2,{m_{H_{0}}}^2,m_W^2)\nonumber\\ 
&+&{e^2({m_{H^{+}}}^2+m_A^2)\over32\pi^2\sin^2\theta_W} 
G_0({m_{H^{+}}}^2,m_A^2,m_W^2)\nonumber\\ 
\end{eqnarray} 
 
\begin{eqnarray} 
\delta Z_{h_0}&=&
{N_Ce^2m_t^2\over32m_W^2\pi^2}\cos^2\alpha\csc^2\beta 
[-B_0({m_{h_0}}^2,m_t^2,m_t^2) 
+(4m_t^2-{m_{h_0}}^2)G_0({m_{h_0}}^2,m_t^2,m_t^2)]\nonumber\\ 
&+&{e^2\cos^2(\alpha-\beta) 
({m_{H^{+}}}^2+{m_{h_{0}}}^2)\over16\pi^2\sin^2\theta_W} 
G_0({m_{h_{0}}}^2,{m_{H^{+}}}^2,m_W^2)\nonumber\\ 
&+&{e^2\sin^2(\alpha-\beta) 
{m_{h_{0}}}^2\over16\pi^2\sin^2\theta_W} 
G_0({m_{h_{0}}}^2,m_W^2,m_W^2)\nonumber\\ 
&+&{e^2\sin^2(\alpha-\beta) 
{m_{h_{0}}}^2\over32\pi^2\sin^2\theta_W\cos^2\theta_W} 
G_0({m_{h_{0}}}^2,m_Z^2,m_Z^2)\nonumber\\ 
&+&{e^2\cos^2(\alpha-\beta) 
(m_A^2+{m_{h_{0}}}^2)\over32\pi^2\sin^2\theta_W\cos^2\theta_W} 
G_0({m_{h_{0}}}^2,m_A^2,m_Z^2)\nonumber\\ 
\end{eqnarray} 
 
Here and below, $B_0$, $C_0$, $C_i$ and $C_{ij}$ is the two-point 
and three-point scalar integrals,  definitions for which can be 
found in Ref. \cite{denner} and $G_0$ is the derivative of $B_0$ 
which is expressed as 
\begin{eqnarray} 
G_0(M^2,m_1^2,m_2^2)={\partial B_0(k^2,m_1^2,m_2^2)\over 
 \partial k^2}|_{k^2=M^2}. 
\end{eqnarray} 
 
\section* {Appendix-B: Self-energy corrections } 
\begin{eqnarray} 
\delta M^{self}_T & = & { N_c e^4 m_{W} \sin(\beta-\alpha) \over 288 \sqrt{2} \pi^2 \hat{s} 
     ( - m_{W}^2 + \hat{s} )^2  \sin\theta_w^4} 
\left[ 6 \hat{s} m_{t}^2 - 2 \hat{s}^2 + 
       3 m_{t}^2 ( m_{t}^2 - 2 \hat{s} )\right. \nonumber \\ 
      & \times & B_0(0,  m_b^2,  m_{t}^2) 
       +  \left. 
       3 ( - m_{t}^4 -\hat{s} m_{t}^2 +2 \hat{s}^2 ) 
        B_0(\hat{s},  m_b^2,  m_{t}^2) \right] 
\end{eqnarray} 
\begin{eqnarray} 
\delta M^{self}_H &=& 
-{e^4m_W\sin^3(\beta-\alpha)\over 
    64\sqrt{2}\pi^2(m_W^2-\hat{s})^2\sin^4\theta_W} 
    \left[ {2\over9}(3 m_H^2-6m_{H^+}^2-\hat{s})+{2m_{H^+}^2\over 
    3}B_0(0,m_{H^+}^2,m_{H^+}^2)\right.\nonumber\\ 
    &+&{(m_{H^+}^2-m_H^2)(m_{H^+}^2-m_H^2-\hat{s})\over3\hat{s}} 
    B_0(0,m_H^2,m_{H^+}^2)\nonumber\\ 
    &+&\left. 
    {[(m_{H^+}-m_H)^2-\hat{s}][\hat{s}-(m_{H^+}+m_H)^2]\over3\hat{s}} 
    B_0(\hat{s},m_H,m_{H^+}^2) \right]\nonumber\\ 
&-&{e^4m_W\cos^2(\beta-\alpha)\sin(\beta-\alpha)\over 
   64\sqrt{2}\pi^2(m_W^2-\hat{s})^2\sin^4\theta_W} 
   \left[{2\over9}(3m_{h_0}^2+6m_{H^+}^2-\hat{s})+{2m_{H^+}^2\over3} 
   B_0(0,m_{H^+}^2,m_{H^+}^2) \right.\nonumber\\ 
   &+&{(m_{H^+}^2-m_{h_0}^2)(m_{H^+}^2-m_{h_0}^2-\hat{s})\over3\hat{s}} 
   B_0(0,m_{H^+}^2,m_{h_0}^2)\nonumber\\ 
   &+&\left.{[(m_{H^+}-m_{h_0})^2-\hat{s}][\hat{s}-(m_{H^+}+m_{h_0})^2]\over3\hat{s}} 
   B_0(\hat{s},m_{H^+}^2,m_{h_0}^2)\right]\nonumber\\ 
&+&{e^4m_W\sin(\beta-\alpha)\over 
   64\sqrt{2}\pi^2(m_W^2-\hat{s})^2\sin^4\theta_W} 
   \left[{2\over9}(6m_A^2+3m_{H^+}^2-\hat{s})+{2m_A^2\over 3} 
   B_0(0,m_A^2,m_A^2)\right.\nonumber\\ 
   &+&{(m_{H^+}^2-m_A^2)(m_{H^+}^2-m_A^2-\hat{s})\over3\hat{s}} 
   B_0(0,m_A^2,m_{H^+}^2)\nonumber\\ 
   &+&\left.{[(m_{H^+}-m_A)^2-\hat{s}][\hat{s}-(m_{H^+}+m_A)^2]\over3\hat{s}} 
   B_0(\hat{s},m_A,m_{H^+}^2)\right]\nonumber\\ 
&-&{e^4m_W\sin(\beta-\alpha)\cos^2(\beta-\alpha)\over 
   64\sqrt{2}\pi^2(m_W^2-\hat{s})^2\sin^4\theta_W} 
   \left[{2\over9}(6m_H^2-\hat{s})+{2m_H^2\over 3} 
   B_0(0,m_H^2,m_H^2)\right.\nonumber\\ 
   &+&({(m_H^2-m_W^2)(m_H^2-m_W^2-\hat{s})\over 3\hat{s}}-{m_W^2\over3}) 
   B_0(0,m_H^2,m_W^2)\nonumber\\ 
   &+&\left.({[(m_H-m_W)^2-\hat{s}][\hat{s}-(m_H+m_W)^2]\over3\hat{s}}-{2m_W^2\over3}) 
   B_0(\hat{s},m_H^2,m_W^2)\right]\nonumber\\ 
&-&{e^4m_W\sin(\beta-\alpha)^3\over 
64\sqrt{2}\pi^2(m_W^2-\hat{s})^2\sin^4\theta_W} 
   \left[{2\over9}(6m_{h_0}^2-\hat{s})+{2m_{h_0}^2\over 3} 
   B_0(0,m_{h_0}^2,m_{h_0}^2) \right.\nonumber\\ 
   &+&({(m_{h_0}^2-m_W^2)(m_{h_0}^2-m_W^2-\hat{s})\over 3\hat{s}}-{m_W^2\over 
   3})B_0(0,m_{h_0}^2,m_W^2)\nonumber\\ 
   &+&\left.({[(m_{h_0}-m_W)^2-\hat{s}][\hat{s}-(m_{h_0}+m_W)^2]\over3\hat{s}}-{2m_W^2\over3}) 
   B_0(\hat{s},m_{h_0}^2,m_W^2)\right]\nonumber\\ 
&+&{e^4m_W\sin(\beta-\alpha)m_H^2\over 
64\sqrt{2}\pi^2(m_W^2-\hat{s})^2\sin^4\theta_W} 
[1+B_0(0,m_H^2,m_H^2)]\nonumber\\ 
&+&{e^4m_W\sin(\beta-\alpha)m_{H^+}^2\over 
64\sqrt{2}\pi^2(m_W^2-\hat{s})^2\sin^4\theta_W} 
[1+B_0(0,m_{H^+}^2,m_{H^+}^2)]\nonumber\\ 
&+&{e^4m_W\sin(\beta-\alpha)m_{h_0}^2\over 
64\sqrt{2}\pi^2(m_W^2-\hat{s})^2\sin^4\theta_W} 
[1+B_0(0,m_{h_0}^2,m_{h_0}^2)]\nonumber\\ 
&+&{e^4m_W\sin(\beta-\alpha)m_A^2\over 
64\sqrt{2}\pi^2(m_W^2-\hat{s})^2\sin^4\theta_W} 
[1+B_0(0,m_A^2,m_A^2)]\nonumber\\ 
\end{eqnarray} 

\section* {Appendix-C: The vertex form factors } 
\begin{eqnarray} 
f_1^{vertex}&=&{-N_c e^4  m_{t}^2 \over {32\,{\sqrt{2}}\,{ 
m_{W}}\,{{\pi }^2}\,\left( {{{  m_{W}}}^2} - \hat{s} \right) \, 
     {{{  \sin\theta_w}}^4}}} 
     \left[ 
      -2   B_0(\hat{s}, m_b^2,  m_{t}^2) 
      \right. 
      \nonumber\\ 
      &+& 
      ( -2  m_{t}^2 - \hat{t} )  C_0( 
        m_{h_0}^2,  m_{W}^2,\hat{s}, m_{t}^2, 
          m_{t}^2,  m_b^2) \nonumber \\ 
          &+& 
         ( -2 m_{W}^2 - \hat{t} )  C_1(  m_{W}^2,\hat{s}, m_{h_0}^2 , 
          m_{t}^2, m_b^2,  m_{t}^2 ) \nonumber\\ 
          &+& 
         ( - m_{h_0}^2 - 2 \hat{t} ) C_2( m_{W}^2,\hat{s},  m_{h_0}^2, 
         m_{t}^2,  m_b^2,  m_{t}^2 ) \nonumber\\ 
         &+& 
       \left. 
       4 C_{00}(m_{W}^2,\hat{s},  m_{h_0}^2 , 
           m_{t}^2,  m_b^2,  m_{t}^2 ) \right], 
      \end{eqnarray} 
     \begin{eqnarray} 
 f_2^{vertex}&=& 
{-N_c \ e^4  m_{t}^2 \over {16\,{\sqrt{2}}\,{  m_{W}}\,{{\pi 
}^2}\,\left( {{{  m_{W}}}^2} - \hat{s} \right) \, 
     {{{  \sin\theta_w}}^4}}} 
  \left[ 
   -  C_0( m_{h_0}^2,m_{W}^2,\hat{s}, 
          m_{t}^2,  m_{t}^2,  m_b^2) 
          \right. 
          \nonumber\\ 
          &-& 
        C_1( m_{W}^2,\hat{s}, m_{h_0}^2, 
          m_{t}^2, m_b^2, m_{t}^2 ) \nonumber\\ 
          & -& 
       3 C_2( m_{W}^2,\hat{s},  m_{h_0}^2 , 
          m_{t}^2,  m_b^2,  m_{t}^2 ) \nonumber \\ 
          &- & 
       2  C_{12}( m_{W}^2,\hat{s},  m_{h_0}^2, 
           m_{t}^2,  m_b^2, m_{t}^2 ) \nonumber \\ 
           &-& 
           \left. 
       2 C_{22}(m_{W}^2,\hat{s},  m_{h_0}^2, 
         m_{t}^2,  m_b^2,  m_{t}^2 ) \right], 
     \end{eqnarray} 
     \begin{eqnarray} 
f_3^{vertex}&=& {-N_c  e^4  m_{t}^2 \over 
   {16\,{\sqrt{2}}\,{  m_{W}}\,{{\pi }^2}\,\left( {{{  m_{W}}}^2} - \hat{s} \right) \, 
     {{{  \sin\theta_w}}^4}}} 
\left[ 
 - C_2(m_{W}^2, \hat{s},  m_{h_0}^2, 
         m_{t}^2,  m_b^2,  m_{t}^2 ) \right. \nonumber \\ 
         &-& 
       2  C_{12}( m_{W}^2,\hat{s},  m_{h_0}^2, 
          m_{t}^2, m_b^2, m_{t}^2 ) \nonumber \\ 
          &-& 
          \left. 
       2  C_{22}( m_{W}^2,\hat{s},  m_{h_0}^2 , 
          m_{t}^2,  m_b^2,  m_{t}^2 ) \right]. 
 \end{eqnarray}

\newpage
\begin{figure}
\epsfxsize=15 cm
\centerline{\epsffile{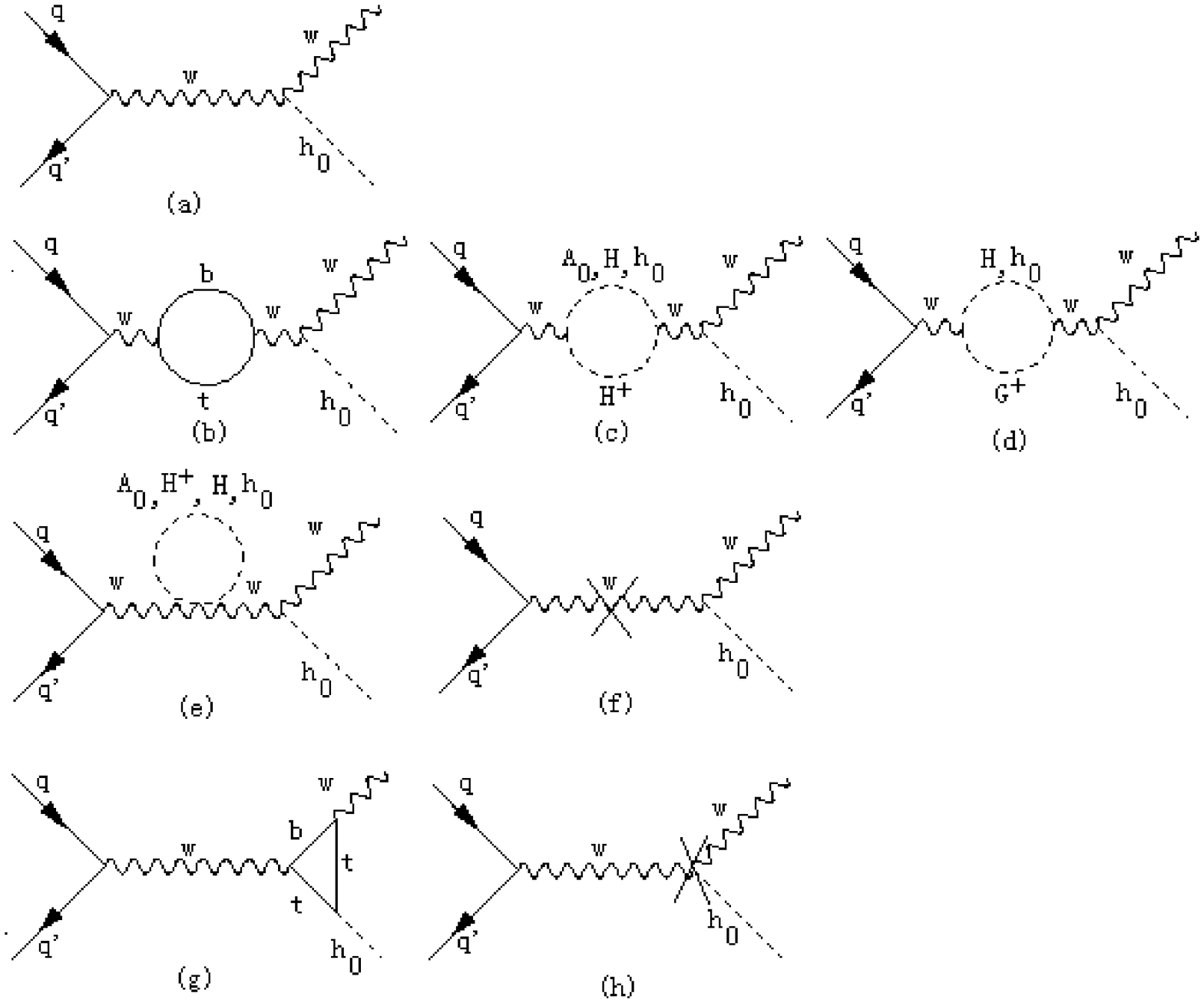}}
\caption[]{}
\label{fig1}
\end{figure}

\begin{figure}
\epsfxsize=15 cm
\centerline{\epsffile{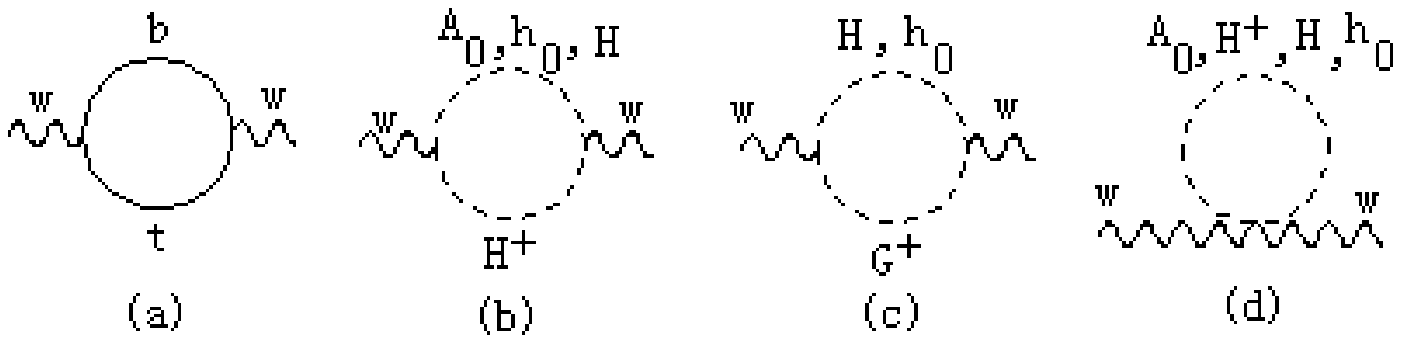}}
\caption[]{}
\label{fig2}
\end{figure}

\begin{figure}
\epsfxsize=15 cm
\centerline{\epsffile{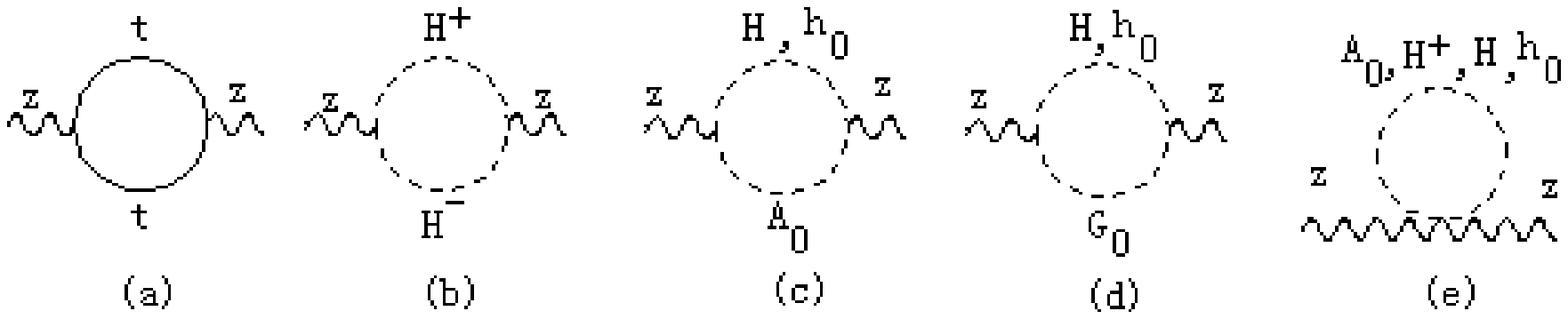}}
\caption[]{}
\label{fig3}
\end{figure}

\begin{figure}
\epsfxsize=15 cm
\centerline{\epsffile{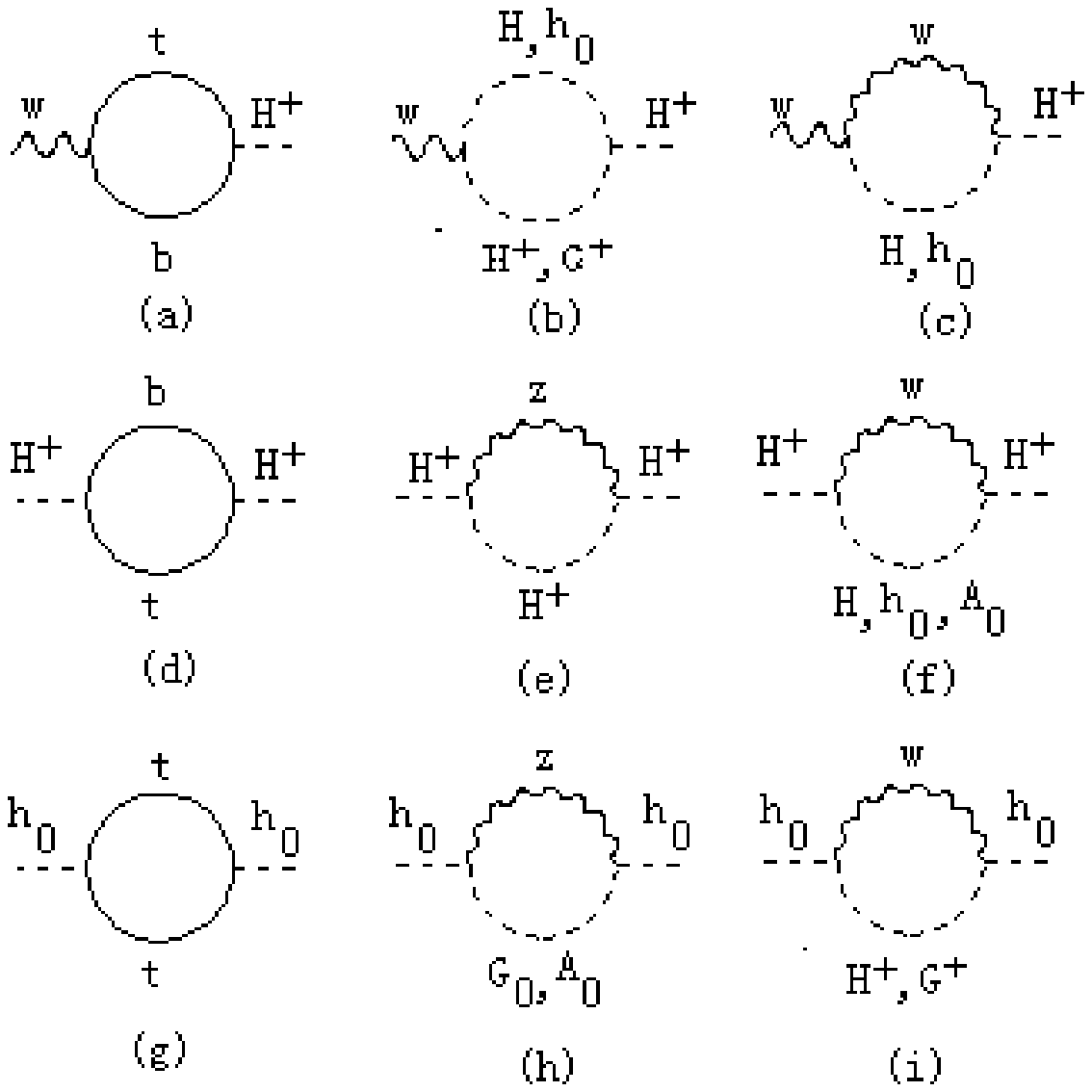}}
\caption[]{}
\label{fig4}
\end{figure}

\begin{figure}
\epsfxsize=15 cm
\centerline{\epsffile{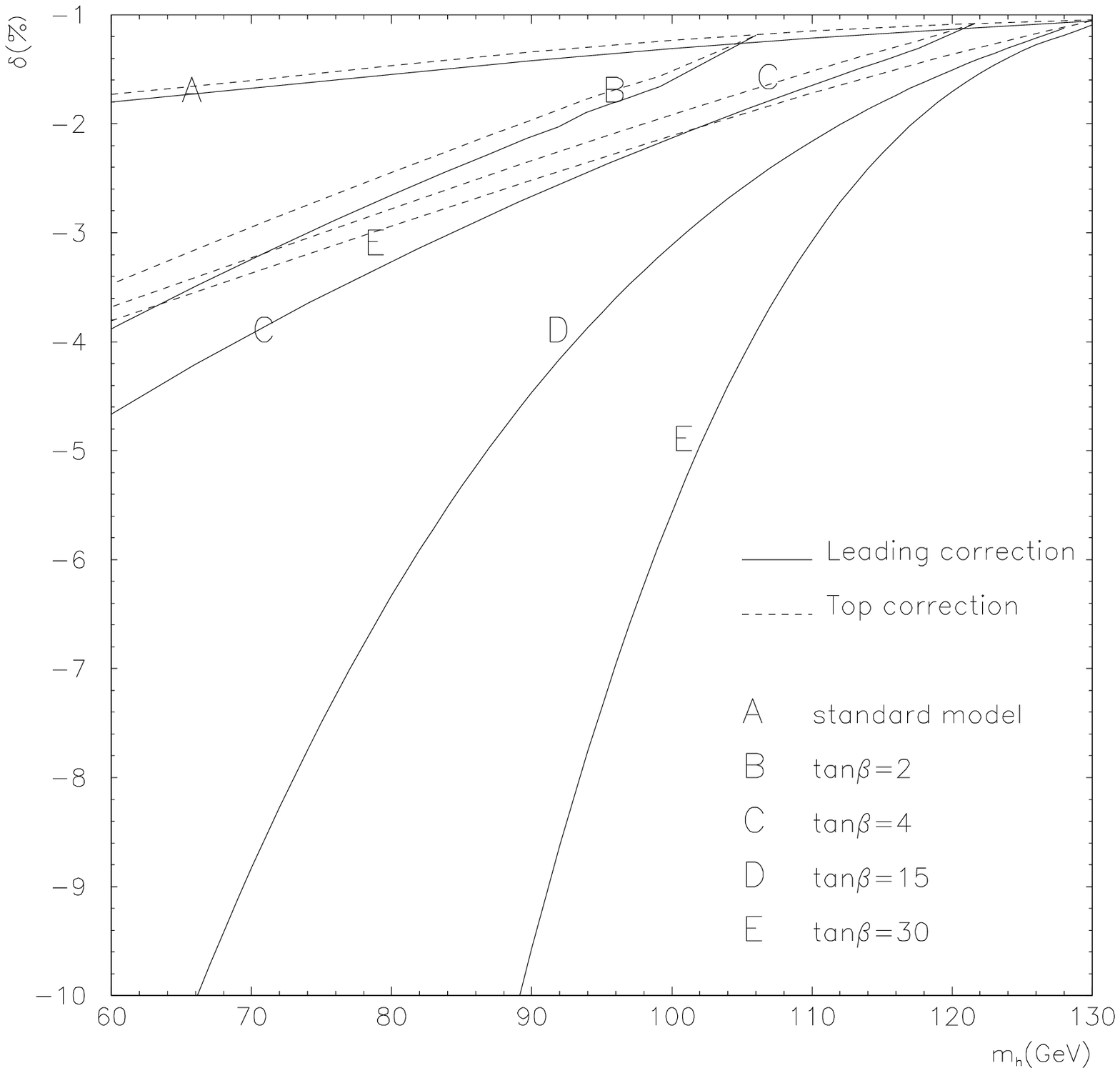}}
\caption[]{Relative corrections $\delta \sigma /\sigma_0 $ 
as a function of the
Higgs boson
mass of the process $q \bar q' \rightarrow W h_0$ 
with $\sqrt{s}=2 TeV$ at Tevatron.}
\label{fig5}
\end{figure}

\begin{figure}
\epsfxsize=15 cm
\centerline{\epsffile{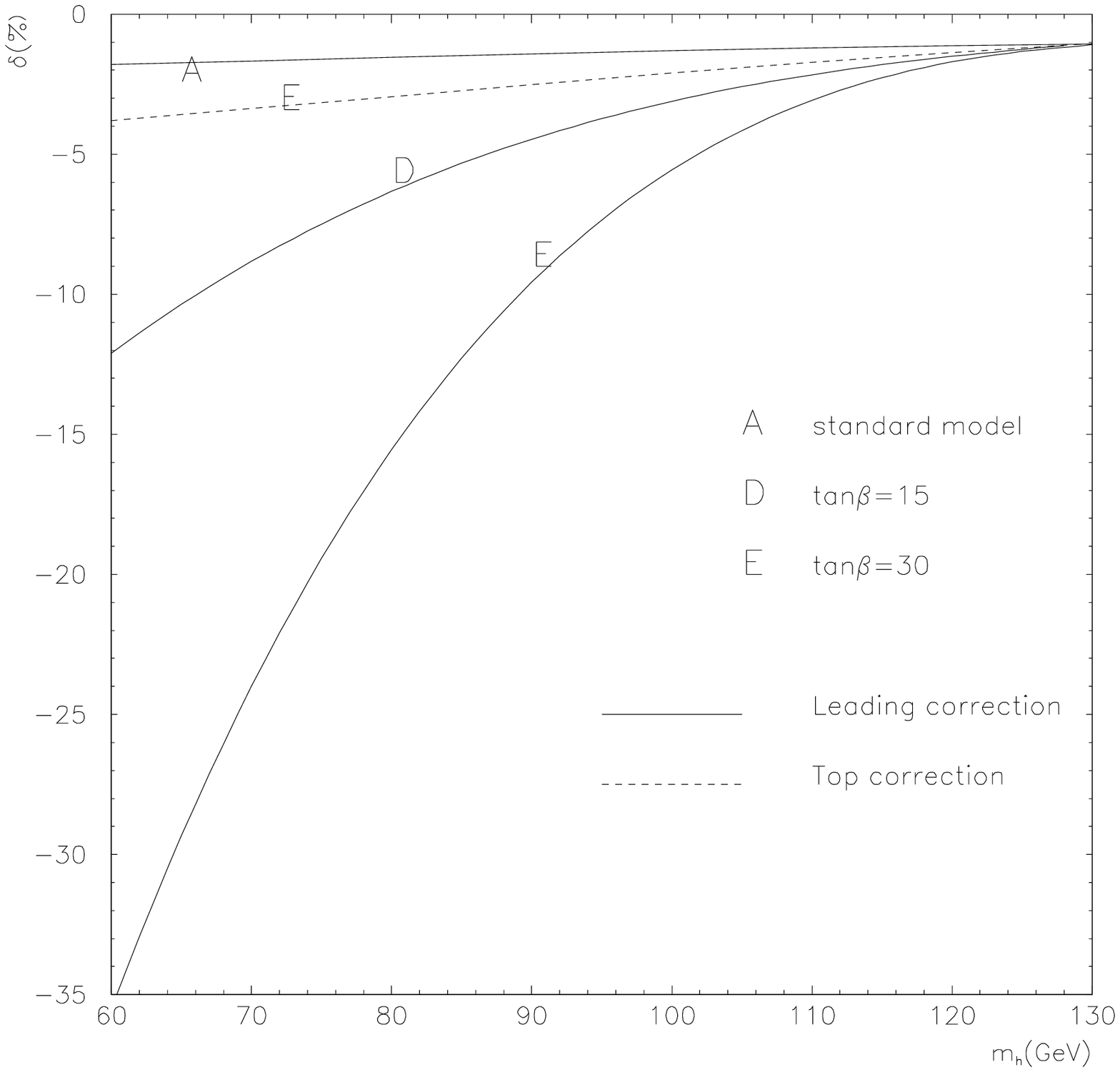}}
\caption[]{Relative corrections $\delta \sigma /\sigma_0 $ 
as a function of the
Higgs boson
mass of the process $q \bar q' \rightarrow W h_0$ 
with $\sqrt{s}=2 TeV$ at Tevatron.}
\label{fig6}
\end{figure}



\begin{thebibliography}{Espinosa} 

\bibitem{a01} 
P. McNamara, {\it ICHEP '98}, Vancouver, July 1998.

\bibitem{a02} 
C. Quigg, FERMILAB-CONF-98/059-T, 
hep-ph/9802320.
 
\bibitem{a03} A. Stange, W. Marciano, and S. Willenbrock, 
{Phys. Rev.} {\bf D49}, 1354 (1994); {\sl ibid.} {\bf D50},4491 (1994); \\
S. Kim, S. Kuhlman, and W.Yao, CDF-ANAL-EXOTIC-PUBLIC-3904,  Oct. 96;\\ 
W.Yao, FERMILAB-CONF-96-383-E,  Jun. 96;\\ 
J. Womersley, D0 Note 3227, Apr. 97;\\ 
S. Parke, FERMILAB-CONF-97/335-T. 
 
\bibitem{a05} 
T. Han and S. Willenbrock, Phys. Lett. {\bf B273}, 167 (1990);\\ 
J. Ohnemus and W.J. Stirling, Phys. Rev. {\bf D47}, 2722 (1993);\\ 
H. Baer, B. Bailay, and J.F. Owens, Phys. Rev. {\bf D47}, 2730 (1993);\\ 
S. Smith and S. Willenbrock, Phys. Rev. {\bf D54}, 6696 (1996). 
 
 
\bibitem{a06} 
H.E. Haber and G.L. Kane, {Phys. Rep.} {\bf 117}, 75(1985);\\ 
J.F. Gunion and H.E. Haber, Nucl. Phys. {\bf B 272}, 1 (1986). 
 
\bibitem{a11} J. Gunion and A. Turski, Phys. Rev. {\bf D39 }, 2701 (1989) and {\bf D40} 
 2333 (1990);\\ 
J.R. Espinosa and M. Quiros, Phys. Lett. {\bf B266}, 389 (1991);\\ 
M. Carena, M. Quiros, and C.E.M. Wagner, Nucl. Phys. {\bf B461}, 407 (1996). 
 
\bibitem{a07} C.S. Li and S.H Zhu, to appear in Phys. Lett. B, hep-ph/9801390. 
 
\bibitem{a08} S. Sirlin, Phys. Rev. {\bf D22}, 971 (1980); W. J. Marciano and A. Sirlin, 
{\sl ibid.} {\bf 22 }, 2695 (1980);{\bf 31}, 213(E) (1985); A. Sirlin and W.J. Marciano, Nucl. Phys. {\bf B189}, 
442 (1981); K.I. Aoki et.al., Prog. Theor. Phys. Suppl. {\bf 73}, 1 (1982). 
 
\bibitem{a09} A. Mendez and A. Pomarol, Phys. Lett. {\bf B279}, 98 (1992). 

\bibitem{CTEQ} H.L. Lai, J. Botts, J. Huston, J.R. Mofin, J.F. Owens, J.W. Qiu, 
W.K. Tung, and H. Weerts, Phys. Rev. {\bf D51}, 4763 (1995). 

\bibitem{add} M. Spiral, CERN-TH/97-68 (hep-ph/9705337).

\bibitem{denner} G. Passarino and M. Veltman, Nucl.Phys. {\bf B33}, 
151 (1979); A. Denner, Fortschr. Phys. {\bf 41} (1993)4. 
 
 
\end{thebibliography}
\end{document}